\documentclass[%
 aip,
 amsmath,amssymb,
preprint,%
]{revtex4-1}

\usepackage{graphicx}
\usepackage{dcolumn}
\usepackage{bm}

\usepackage[utf8]{inputenc}
\usepackage[T1]{fontenc}
\usepackage{mathptmx}
\usepackage{etoolbox}

\usepackage{chemformula}
\usepackage{siunitx}

\makeatletter
\def\@email#1#2{%
 \endgroup
 \patchcmd{\titleblock@produce}
  {\frontmatter@RRAPformat}
  {\frontmatter@RRAPformat{\produce@RRAP{*#1\href{mailto:#2}{#2}}}\frontmatter@RRAPformat}
  {}{}
}%
\makeatother
\begin{document}

\preprint{AIP/123-QED}

\title[Band gap analysis and carrier localization in cation-disordered \ch{ZnGeN2}]{Band gap analysis and carrier localization in cation-disordered \ch{ZnGeN2}}
\author{Jacob J. Cordell}
\email{cordell@mines.edu}
\affiliation
{Department of Mechanical Engineering, Colorado School of Mines, Golden, CO, USA}
\affiliation
{Materials Science Center, National Renewable Energy Laboratory, Golden, CO, USA}
\author{Garritt J. Tucker}
\affiliation
{Department of Mechanical Engineering, Colorado School of Mines, Golden, CO, USA}
\author{Adele Tamboli}
\affiliation
{Materials Science Center, National Renewable Energy Laboratory, Golden, CO, USA}
\author{Stephan Lany}
\email{stephan.lany@nrel.gov}
\affiliation
{Materials Science Center, National Renewable Energy Laboratory, Golden, CO, USA}

\date{\today}

\begin{abstract}
The band gap of \ch{ZnGeN2} changes with the degree of cation site disorder and is sought in light emitting diodes for emission at green to amber wavelengths. By combining the perspectives of carrier localization and defect states, we analyze the impact of different degrees of disorder on electronic properties in \ch{ZnGeN2}, addressing a gap in current studies which largely focus on dilute or fully disordered systems. The present study demonstrates changes in the density of states and localization of carriers in \ch{ZnGeN2} calculated using band gap-corrected density functional theory and hybrid calculations on partially disordered supercells generated using the Monte Carlo method. We use localization and density of states to discuss the ill-defined nature of a band gap in a disordered material and identify site disorder and its impact on structure as a mechanism controlling electronic properties and potential device performance. Decreasing the order parameter results in a large reduction of the band gap. The reduction in band gap is due in part to isolated, localized states that form above the valence band continuum associated with nitrogen coordinated by more zinc than germanium. The prevalence of defect states in all but the perfectly ordered structure creates challenges for incorporating disordered \ch{ZnGeN2} into optical devices, but the localization associated with these defects provides insight into mechanisms of electron/hole recombination in the material.
\end{abstract}


\maketitle 

\section{Introduction}
\label{sec:intro}

Site disorder, the replacement of chemical species on a fixed crystallographic lattice, has recently grown in interest across semiconductor research areas as a means to control optoelectronic properties. While site disorder–referred to from here on simply as disorder or degree of order–has notably been studied as a mechanism for managing properties in chalcogenide transistor and solar cell materials for some time, \cite{farges1997ti,ahmed1997deposition,martins2007role,katagiri2008enhanced,chen2009crystal} its application to such a vast array of ternary and multinary nitrides and phosphides is a more recent development. \cite{schnepf_ACS_2020,ogura2021electronic} Insight from broader comparisons of \ch{II-IV-N2} materials has identified relationships between cation species, structural distortion and electronic structure due to this disorder \cite{kute2021cation} and in some systems site disorder has been investigated as a means of lowering band gap energies to ideal ranges for targeted applications. In \ch{ZnGeN2}, cation disorder is sought to reduce the band gap from the calculated \cite{melamed2020combinatorial,Punya2011,adamski2017} and measured \cite{kikkawa1999rf,Du2008,osinsky2000new,Viennois2001} 3.0-3.6 eV to the 2.1-2.5 eV range desired for amber to green wavelengths in a light emitting diode (LED), often referred to as the green gap.\cite{phillips2006basic} Disordered \ch{ZnGeN2}, which is lattice matched to GaN, may be desirable as a replacement for high In content In$_x$Ga$_{1-x}$N, which suffers from a miscibility gap and large lattice mismatch with GaN in heterostructure devices. \cite{ssl2019lighting,tellekamp2020heteroepitaxial,reeber2000,kikkawa1999rf} 

\begin{figure}[ht]
 \begin{center}
 {\includegraphics[scale=0.38]{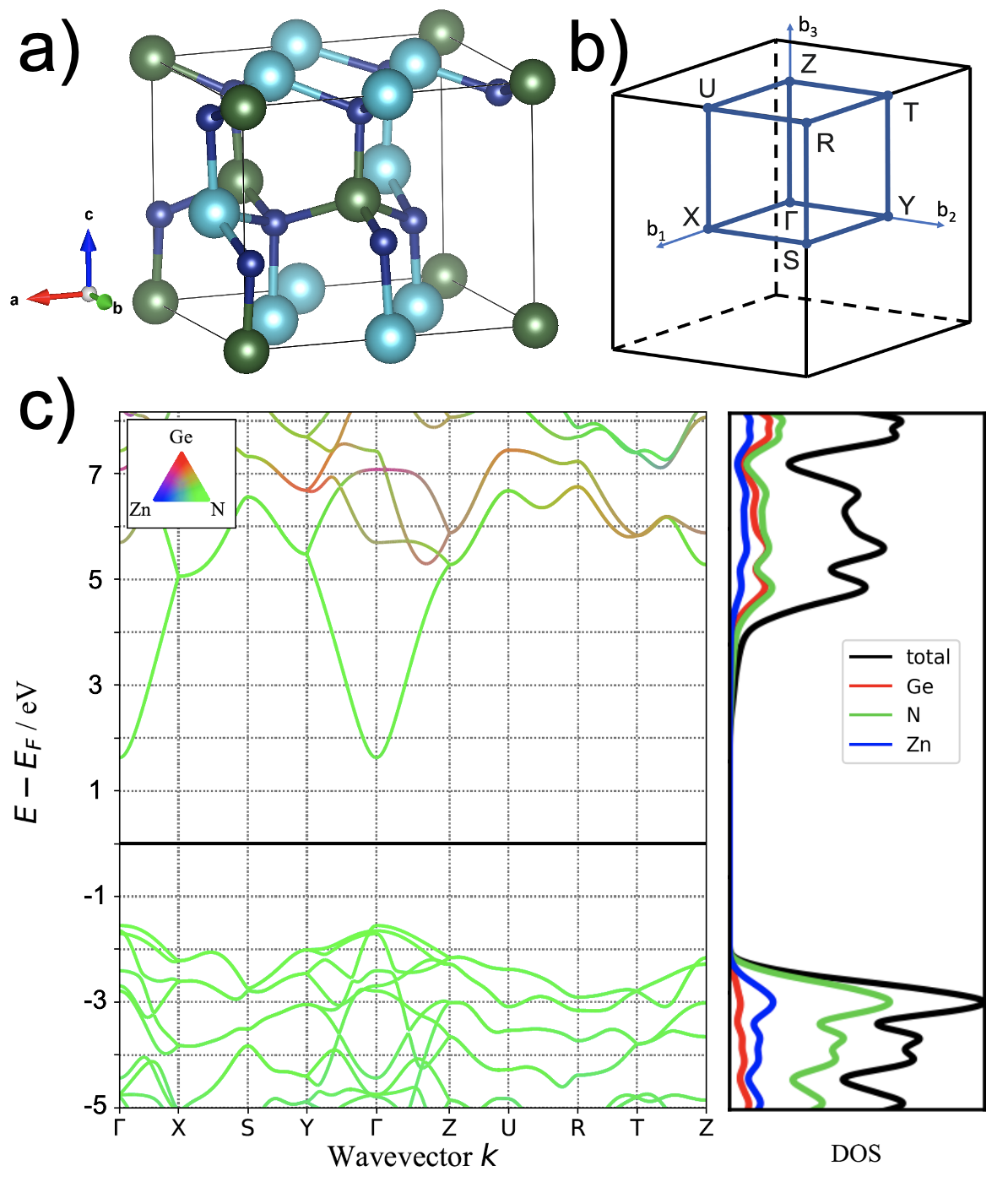}}
 \caption{a) Ordered ground state crystal structure of \ch{ZnGeN2} b) reciprocal space map of \ch{ZnGeN2} ($b_3 > b_2 > b_1$) c) band structure and DOS of ordered \ch{ZnGeN2}. }
 \label{fig:intro}
 \end{center}
\end{figure}

Because disorder adds nuance to how a band gap is measured and calculated, when the term `band gap' is used in this work, we refer to the energy difference between the highest occupied and lowest unoccupied molecular orbitals (HOMO and LUMO) unless specified otherwise. However, this energy difference is not the only viable definition as will be discussed throughout this Article. To investigate the impact of disorder on the band gap of \ch{ZnGeN2}, we utilize disordered structures in large supercells of 1,024 atoms \cite{cordell_2021_probing}. These structure models incorporate site disorder consisting of cation antisite pairs, which numerous defect studies have highlighted as the dominant native defects in \ch{ZnGeN2}. \cite{Skachkov2016,Skachkov2016_doping,adamski2017,Skachkov2017,lyu_band_2019,melamed2020combinatorial,Skachkov2020,haseman2020} In contrast to a dilute defect model, site disorder accounts for the interaction of $Zn_{Ge}$ and $Ge_{Zn}$ present in high concentrations representative of materials grown under non-equilibrium conditions. This study separates the impact of site disorder explicitly from stoichiometry, non-native defects and crystalline quality, all known to further influence optical and electronic properties of interest. To illustrate the ordered system, Figure \ref{fig:intro} provides the crystal structure, reciprocal space map and band structure of \ch{ZnGeN2}.

In a dilute defect picture, defects do not interact and additional occupied or unoccupied states are viewed as defect states within an otherwise unchanged band gap. Historically, the theoretical discussion of differentiating band gaps and defect states has been held in this context of dilute point defects \cite{noda2000trapping,lany2007dopability,freysoldt2014first} or in fully random systems \cite{wei1990band,veal2015band,xiang2008strain}, but misses systems with intermediate degrees of order with a few notable exceptions \cite{chan2010bridging,liu2016special}. 

In materials with both dilute and non-dilute defects, Urbach energy \cite{Urbach1953} describes how the optical absorption of a semiconductor tails off exponentially \cite{Soukoulis1984,John1986,Sa-Yakant1987} at energies below the band gap due to transitions from within bands to defect states in the energy gap and at even lower energies directly between defect states in the gap. \cite{Shimakawa2006ch3,adachi2012optical,Sharma2017} Urbach tails are evident in Tauc \cite{tauc1966optical} analyses of thin films as well as in bulk systems, where variations of the Kubelka-Munk method \cite{kubelka1931article,kubelka1948new} are often used to interpret band gaps. These bulk and film methods frequently vary in interpretation of an optical band gap based on differences in their assumptions. \cite{saenz2016optical,dolgonos2016direct,makula2018correctly}

The difficulty in properly defining a band gap stems to a large extent from the fact that the band gap is used as a scalar metric to address a multitude of related but distinct phenomena and questions, either in experimental measurements or theoretical computation, and in various fields of research. Fundamentally, the band gap is the difference between ionization potential (electron removal energy) and electron affinity (electron addition energy). As such, it is not an optical or even excited-state property. However, most experimental approaches for band gap measurements are based on optical spectroscopy, as mentioned above. In such approaches, it is difficult to account for nontrivial physical mechanisms that modify the shape of the spectra from which the band gap value is deduced. For example, calculations using the Bethe-Salpeter equation (e.g., Ref.~\citenum{laskowski1005}) show that excitonic effects (electron-hole interaction) tend to redshift the dielectric response above the absorption threshold compared to the independent particle approximation, and enable sub-band gap excitations (exciton binding energy). Similarly, the variation of oscillator strength resulting from wavefunction symmetries (direct vs indirect, allowed vs forbidden transitions) is often not precisely known, but can affect the spectra in ways that are not fully captured by model parameters used, e.g. in Tauc analysis. Furthermore, there is a fundamental difference between optical transition energies in absorption and emission, i.e., the Stokes shift \cite{odonnell1999}, which is non-radiatively converted to heat. These effects add significant uncertainties to band gap determination in all but the most thoroughly characterized systems (e.g., \ch{GaAs} \cite{nam1976}, \ch{Cu2O} \cite{kazimierczuk2014giant}, \ch{ZnO} \cite{manjon2003}, \ch{GaN} \cite{monemar1974}). These uncertainties are further exacerbated in disordered materials, where one must make additional assumptions or define models to discriminate between defect and continuum states.

This work addresses these challenges from the perspective of large-scale supercell electronic structure calculations in disordered \ch{ZnGeN2}. Here, we investigate the consequences of disorder in \ch{ZnGeN2} due to non-equilibrium synthesis on electronic structure. We use non-self-consistent hybrid functional calculations to enable analysis of the density of states (DOS) and carrier localization as a function of long-range order (LRO) and short-range order (SRO). The band gap of 3.5 eV of ordered \ch{ZnGeN2} decreases with increasing degree of disorder and eventually closes for strongly disordered configurations. Calculated inverse participation ratios (IPR) allow us to assess the localization of states in this range of disordered \ch{ZnGeN2} and discuss how localization impacts our interpretation of a band gap as well as device characteristics. By comparing the DOS of \ch{ZnGeN2} structures from band gap corrected calculations in 1,024 atom cells, we analyze the effect of disorder on the band gap of the system.

\section{Disordered atomic structure models}
\label{sec:models}

This Article builds on results from previous work \cite{cordell_2021_probing} in which disordered \ch{ZnGeN2} structures were generated using Monte Carlo (MC) simulations, providing atomic structure models with systematic variation of the order parameters across the order-disorder transition. The degree of disorder is controlled by an effective temperature describing the site ordering of a cation configuration within a crystalline system. This model includes the configurational entropy contribution to the free energy of the system, but excludes factors such as decomposition reactions which dominate at higher actual synthesis or process temperatures. Thus, the effective temperature provides a link to map site disorder between MC simulations and non-equilibrium synthesis. \cite{cordell_2021_probing} We focus in this work on four effective temperatures representing four separate regimes of ordering. 2,000 K and 2,500 K structures include the ground state, ordered configuration as well as mostly ordered structures with few antisites per cell. 3,000 K structures are disordered, but not random and 5,000 K structures are highly disordered but still not random. The level of disorder between 3,000 K and 5,000 K is best understood through differences in electronic properties as discussed later in this Article. Truly random configurations are not realized below effective temperatures of approximately 400,000 K \cite{cordell_2021_probing}.

To relate DOS, IPR and ordering, we employ the fraction of nitrogen coordinated by exactly two Ge and two Zn (Zn$_2$Ge$_2$ motif fraction) as a measure of SRO, as well as the Bragg-Williams LRO parameter, $\eta$:
\begin{equation}
    \eta = r_{Zn} + r_{Ge} -1
\end{equation}

\noindent
where $r_{Zn}$ ($r_{Ge}$) is the fraction of \ch{Zn} (\ch{Ge}) on  \ch{Zn} (\ch{Ge}) ground state sites. \cite{Bragg1934,Bragg1935} 

Both LRO and SRO parameters–measures of Wyckoff site occupancy and nitrogen coordination, respectively–indicate full ordering at low effective temperatures. Both parameters slightly decrease from their ordered values of one when individual defect pairs (site exchange of Zn and Ge) are introduced in the supercells with increasing effective temperature. LRO and SRO parameters then drop abruptly at the transition of 2,525K. Above the transition temperature, the order parameters taper from small values to their fully disordered extremes at an infinite effective temperature, 0 for LRO and 0.375 for SRO. The transition in order parameters covers a wider range of accessible SRO than LRO parameters, but the transition occurs in both length scales simultaneously (LRO and SRO are strongly coupled). This behavior contrasts with the \ch{ZnSnN2:ZnO} system, \cite{pan2020perfect} where SRO can exist without LRO.

\section{Disorder and Density of States}
\label{sec:results}

Paying special attention to the role of localized states in determining the value of the band gap, we assess what a band gap means in the context of disordered solids. We use the IPR here as a measure of the localization of a state at a given energy as shown in Equation \ref{eqn:ipr}, where the IPR indicates that a given state at a given energy is localized on average on one out of IPR atoms. An IPR of 1 indicates perfect delocalization and a value of 1,024 indicates exclusive localization on a single atom within the supercell.
 
\begin{equation} \label{eqn:ipr}
     IPR(E) = \frac{N_A\sum_i p_i (E)^2}{[\sum_i p_i(E)]^2}
\end{equation}

\noindent
$N_A$ is the number of atoms in a supercell and $p_i (E)$, the local density of states (LDOS) projected on each atom $i$ as a function of energy $E$.

\begin{figure}[ht]
 \begin{center}
 {\includegraphics[scale=0.18]{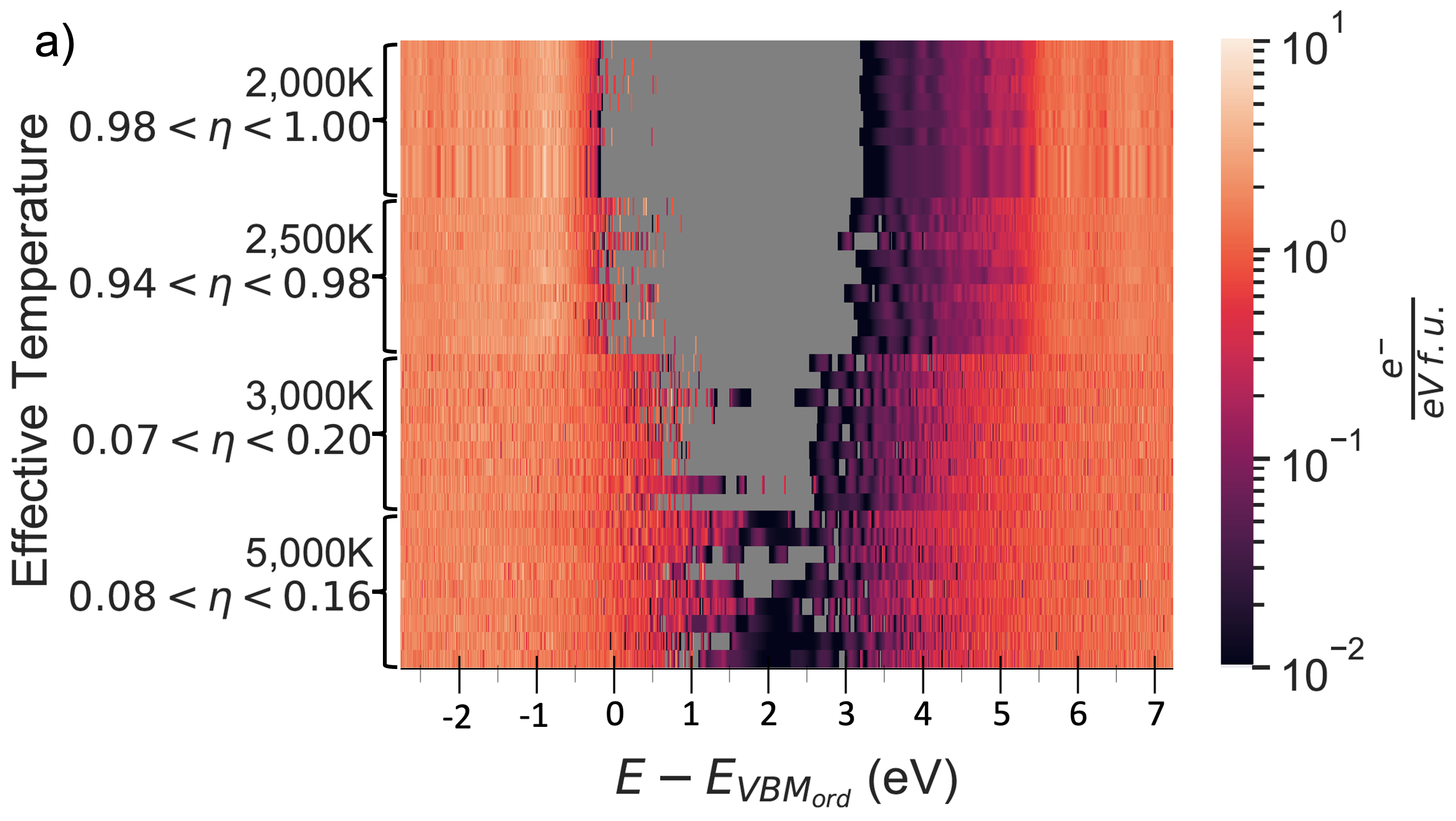}}
 {\includegraphics[scale=0.18]{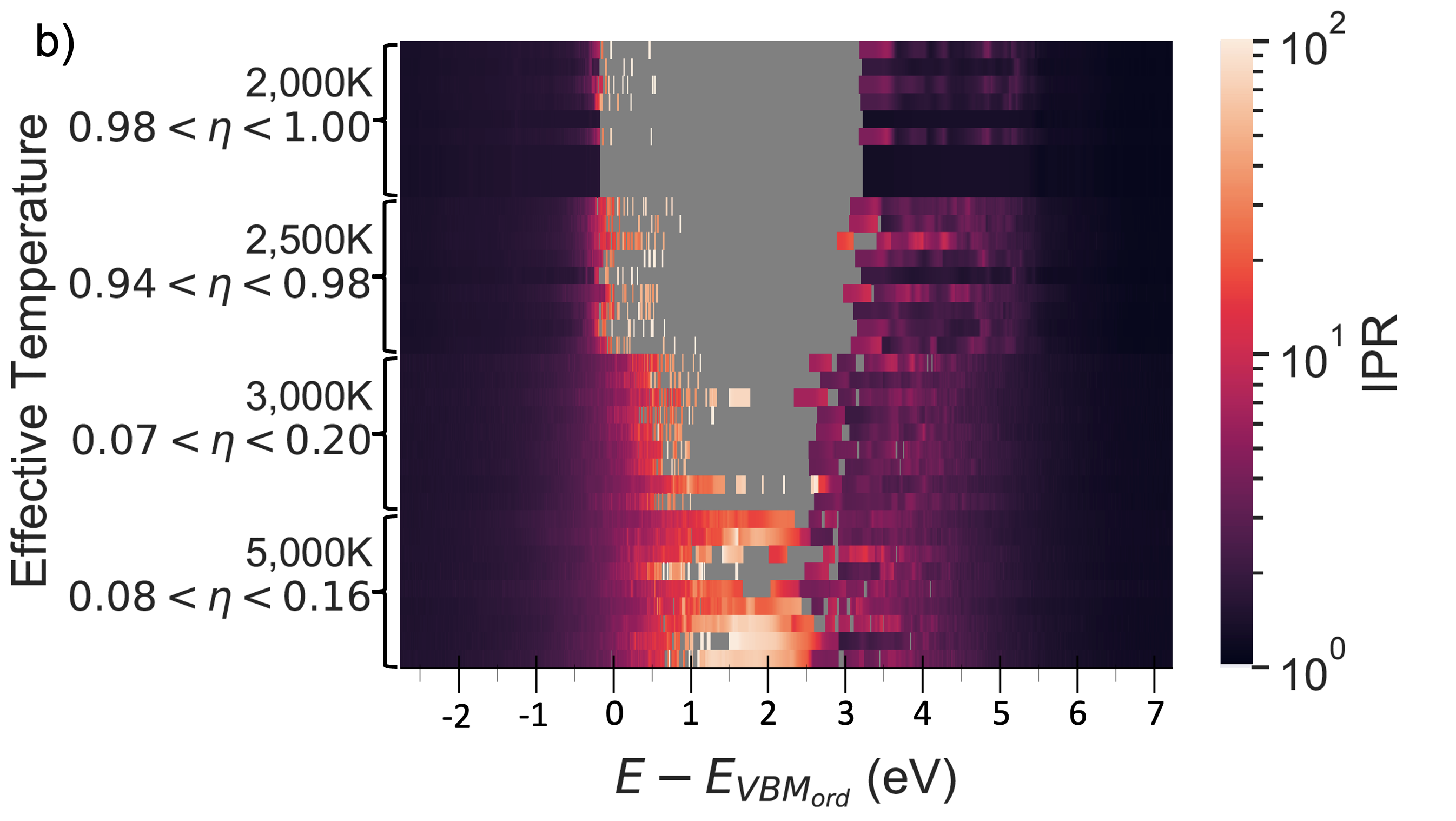}}
 \caption{a) Total DOS and b) IPR of \ch{ZnGeN2} configurations with four distinct effective temperatures.}
 \label{fig:dos}
 \end{center}
\end{figure}

Figure \ref{fig:dos}a) shows the total DOS of 36 \ch{ZnGeN2} configurations evenly grouped by their effective temperature and corresponding range of the LRO parameter. To access the effect of disorder on valence (occupied) and conduction (unoccupied) band states individually, we determined the potential alignment of the disordered structures relative to the ground state (see Section \ref{sec:methods}). Using a 10 meV increment in data points, defect states appear in the DOS of disordered structures represented by allowed (non-gray) bands separated from the band edges by forbidden (gray) states. Up to $T_{eff}$ = 2,000 K, the MC simulation largely retains the ordered ground state structure, but some $Zn_{Ge}$ and $Ge_{Zn}$ antisite configurations start to develop. Between 2,000 K and 2,500 K, the concentration of antisite defects increases with a concomitant decrease in the average band gap of 0.7 eV. Just above 2,500 K, the system undergoes an order-disorder phase transition \cite{cordell_2021_probing}, assuming a state with both long- and short-range disorder. It is important to note, however, that the system retains a significant degree of non-random LRO and SRO up to much higher effective temperatures. As seen in Figure \ref{fig:dos}, comparing $T_{eff}$ = 2,500 K and 3,000 K, the phase transition is accompanied by a large reduction of the order parameter $\eta$ and an additional band gap reduction of about 1.1 eV. The average band gap then decreases by 1.0 eV from 3,000 K to 5,000 K as the system tends toward metallic for mostly disordered structures.

The decrease in band gap with disorder comes from movement in both the conduction band minimum (CBM) and valence band maximum (VBM); replacing a single pair of cations in the ground state structure with an antisite pair raises the Fermi level by 200-500 meV depending on the proximity of the pair. The Fermi level in this context is taken as the midpoint between the energy levels of the highest occupied and lowest unoccupied states and lies just below 2 eV on the energy scale of Figure \ref{fig:dos}. Further decreasing the degree of order does not significantly impact the Fermi level beyond the initial shift until the band gap effectively closes. Without contributions from non-native, non-antisite defects and stoichiometry, site disorder alone drastically changes the band gap of \ch{ZnGeN2} over a 3.5 eV range.

Figure \ref{fig:dos}b) provides the corresponding IPR of the DOS from Figure \ref{fig:dos}a) allowing a look at the localization of states. IPR is undefined where the DOS is zero (in the band gap). Like the DOS, IPR are discretized with a step size of 10 meV. The scaling of the color bar representing IPR in Figure \ref{fig:dos}b) highlights the most localized states, the movement of which can be tracked across the band gap (gray region) with changing order parameter. These localized states remain relatively constant in energy relative to the VBM but increase in quantity and density with decreasing degree of order. The localized, mid-gap states indicate a high probability of non-radiative recombination centers in disordered \ch{ZnGeN2}, \cite{street1977non,kawakami2001radiative} which coincides with lower conversion efficiencies in LEDs.

While configurations with both small and large (disordered) fractions of antisites have states with relatively high IPRs, one identifying factor for interpreting band gaps is the range of these localized states in energy.  Individual antisite pairs result in defect states that can clearly be associated with either the conduction or valence band edge; however, this clarity is lost as the gap closes and high IPR states spread throughout the gap. At higher effective temperatures represented in Figure \ref{fig:dos}b), for instance, states with an IPR above 50 are all occupied even though some appear high in energy, adjacent to less localized, unoccupied states; this indicates a closed energy gap. The Fermi levels for metallic configurations in Figure \ref{fig:dos}b) fall between 2.5 eV and 3 eV on the provided energy scale. The high IPR values in the gap, close to the valence band edge for dilute defect structures are strongly correlated with the \ch{Zn3Ge1} motif centered on nitrogen whereas the \ch{Zn1Ge3} and \ch{Zn2Ge2} motifs contribute significantly less to the IPR. \ch{Zn4Ge0} and \ch{Zn0Ge4} motifs only appear in very limited cases below 3,000 K. At and above 2,500 K, some conduction band states separate from the continuum in energy. The highest contributor to the IPR in these bands is mostly Zn with a few instances of Ge contributing more than Zn and N. The participation of the cations independent of their first and second shell coordination contrasts separated valence bands where nitrogen and its first shell coordination play the largest role. The IPR in these separated conduction band states reaches a maximum of 20, significantly less than the maximum of 200 for elevated valence band states.

\begin{figure}[ht]
 \begin{center}
 {\includegraphics[scale=0.22]{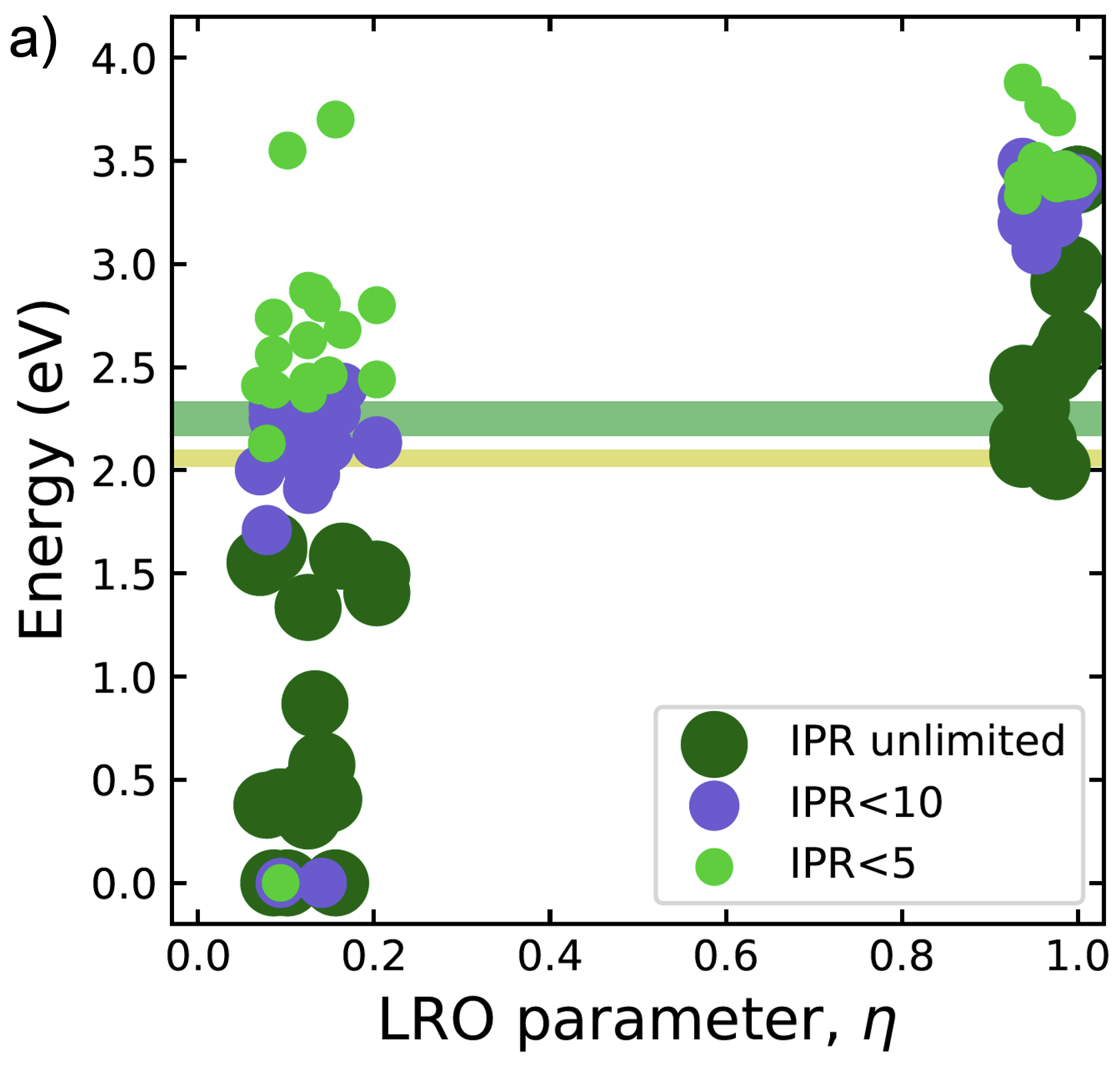}}
 {\includegraphics[scale=0.22]{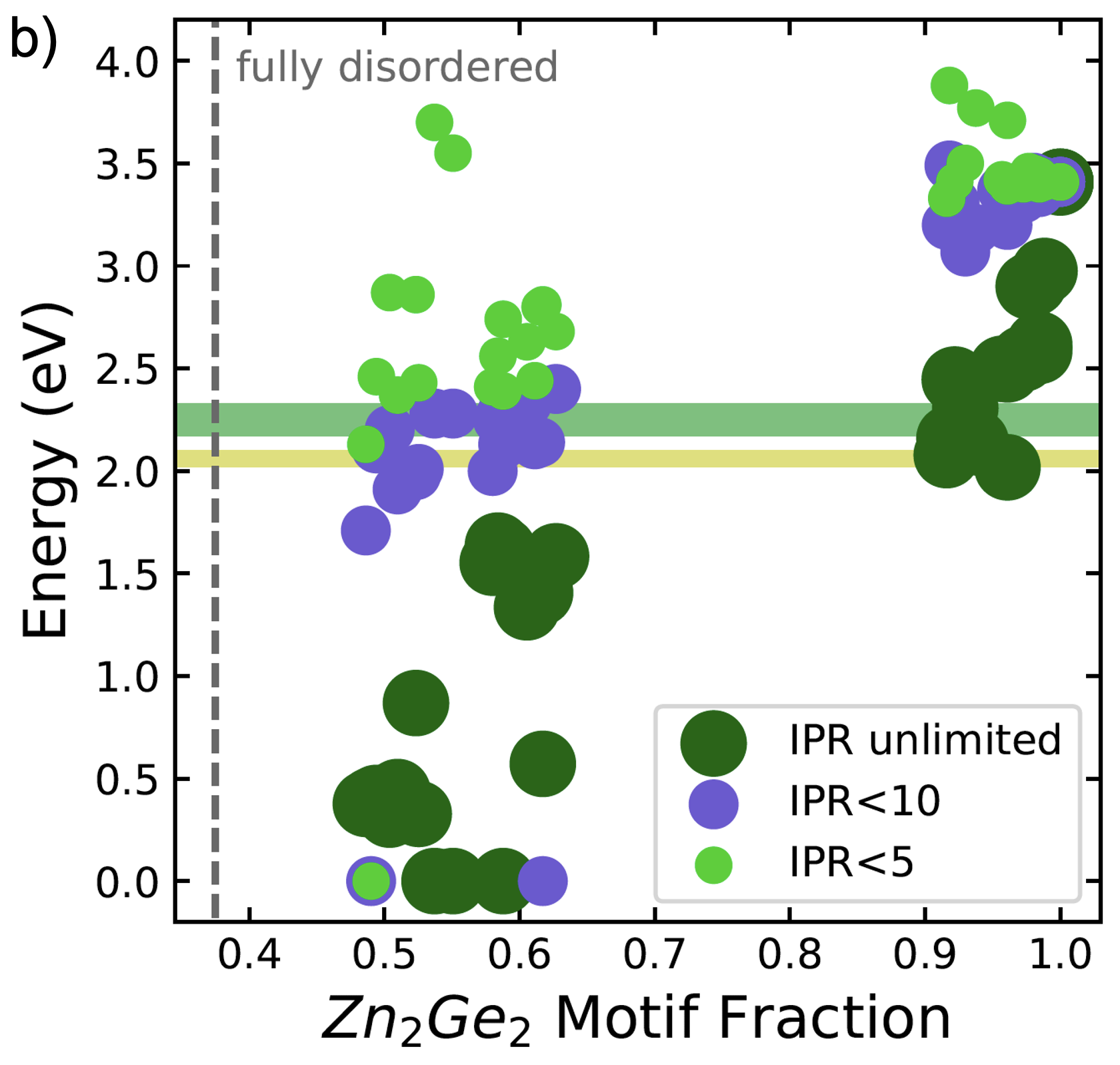}}
 \caption{Band gap energy as a function of a) long-range order and b) short-range order for three interpretations of the band gap. IPR unlimited: difference in energy between highest occupied and lowest unoccupied molecular orbitals. IPR $<$ 10 (5): States with IPR $>$ 10 (5) are excluded from the band gap determination.}
 \label{fig:gaps}
 \end{center}
\end{figure}

Until this point, we have treated the band gap of a system simply as the difference in energy between the highest occupied and lowest unoccupied states as this definition is typically effective in ordered systems. This definition often results in the inclusion of defect states as part of the band continuum when in reality, these isolated states hinder electronic transport and can be considered defect states inside the band gap. In this context, ``bands'' are typically assumed to consist of delocalized states, but Figure \ref{fig:dos}b) shows some states in disordered solids are in fact highly localized. The IPR of a system is large for defect states and shrinks for states in the band continuum thus providing a measure of the extent to which a state acts as a defect. Using this information, we define an alternative band gap where only states below a certain IPR threshold are allowed when taking the highest occupied and lowest unoccupied states. A high IPR cutoff (or no cutoff in the case of unlimited IPR) yields the smallest gaps, whereas a low cutoff considers only continuous conduction and valence band states and results in the largest gaps. In the ground state structure, the IPR at the VBM and CBM are 1.54 and 1.29, respectively. IPR limits of 5 and 10 are used in Figure \ref{fig:gaps} to give mid- and low- IPR cutoff examples. Providing too low of a cutoff excludes valid continuum states giving unphysical band gaps larger than that of the ground state structure. In Figure \ref{fig:gaps}, these band gaps are plotted over the full range of the LRO ($0<\eta<1$) and SRO ($0.375<Zn_2Ge_2<1$) parameters and amber/green bands are drawn at their relevant energies for comparison of calculated band gaps in this region.

Interpreting band gaps both as-calculated and after excluding states with an IPR above a certain threshold allows us to determine the contribution of those localized states to the electronic structure as well as where those states cluster in energy. In Figure \ref{fig:gaps}, the traditional, unlimited IPR definition shows how small fractions of defects drastically reduce the band gap, while ignoring highly localized states shows that this significant change is largely–but not exclusively–due to these isolated defects. In the band gap interpretations that do not consider highly localized states (e.g., IPR$<$10 and IPR$<$5), the band gap still decreases by roughly 2 eV, but this reduction occurs through continuous bands in energy in disordered configurations rather than through defect states in structures with near-perfect ordering. 

Taking band gaps as the difference between highest occupied and lowest unoccupied states yields a change from 3.5 eV to 2.0 eV with only a drop in LRO parameter from $\eta$=1.00 to $\eta$=0.94. For low order parameters, the difference between the unlimited gaps and gaps excluding states with IPR$>$5 is again significant. The IPR limitation places the gap in the amber/green region of the visible spectrum with some trend toward higher band gap with higher SRO parameter per Figure \ref{fig:gaps}b). For largely disordered structures, this sizable transition creates very small band gaps less than 1.6 eV for $\eta\leq0.20$, a much larger change in band gap with ordering than predicted for the more researched \ch{Cu2ZnSnS} (CZTS) system. \cite{scragg2016cu,Zawadzki2015} When these structures are fully random (i.e., at infinite effective temperature), the band gaps consistently drop to zero for the unlimited definition and are undefined for the cases with limited IPR.

Though the supercell size of structures used for the present analysis is large for typical DFT calculations (and especially so for band gap corrected electronic structure calculations), it is still limited in capturing the statistics of configurational disorder, particularly in the dilute defect limit (low effective temperature). The localization of these states in structures with small fractions of defects and the impact of the defects' spatial proximity were studied by Skachkov \emph{et al.} \cite{Skachkov2016}. These mid-gap states isolated in energy are generally accepted as detrimental to optoelectronic properties by decreasing the quantity of carriers collected, reducing the lifetimes of those carriers or inhibiting radiation of a photon. \cite{park2018point,meneghini2014characterization} However, at higher defect concentrations, where defect bands widen in energy as in largely disordered supercells in Figure \ref{fig:dos}b), conflicting theories exist as to the impact of defect density on the relative rate of non-radiative recombination.

In one theory, Luque \emph{et al.} directly connect non-radiative Shockley-Read-Hall recombination to the localization caused by a low density or irregularity of impurities within a lattice, but see a reduction in non-radiative recombination as defect density increases above a certain threshold \cite{luque2006intermediate}. In this theory, lower densities of defects correspond to more spatially isolated and therefore localized defects and spatially connected states exhibit more benign electronic properties \cite{luque2012understanding}. However, gap states in \ref{fig:dos}b) show a comparable maximum IPR for every structure other than the ground state, independent of the degree of disorder of those configurations. These comparable degrees of localization independent of defect density align better with prevalent studies in the \ch{InGaN2} system. In \ch{InGaN2} and similar III-V alloys, higher defect densities and deep gap states cause higher rates of non-radiative recombination \cite{yang2016non,lee2017point,shabunina2013extended}. Based on the high degree of localization in disordered configurations, this latter theory of higher defect densities negatively impacting radiative recombination applies to \ch{ZnGeN2} as well.

In order to address the non-radiative energy loss in disordered \ch{ZnGeN2}, we performed calculations of the electron and hole capture processes. These calculations require self-consistent hybrid functional calculations to overcome the delocalization error of standard DFT \cite{mori2008localization,lany2011predicting}. For this purpose, we used a 128 atom cell with one distant $Zn_{Ge}-Ge_{Zn}$ anti-site pair as an exemplary configuration. The non-radiative energy loss (Stokes shift) during optical recombination corresponds to the atomic relaxation energies after carrier capture and recombination, which can be represented in a configuration coordinate (Franck-Condon) diagram \cite{peng2012semiconducting}. For the hole trapping, we obtain a relaxation energy of -0.29 eV after localization at a N site adjacent to the ZnGe anti-site, and -0.34 eV for the relaxation back to the initial state after recombination with an electron. The sum of these energies is converted to heat and reduces the photon energy relative to the HOMO-LUMO gap. For this specific anti-site pair configuration, the electron localization at the $Ge_{Zn}$ site is energetically unfavorable by +0.38 eV compared to the delocalized conduction band like state, making electron capture unlikely. It is possible that in a more strongly disordered state, electron capture becomes exothermic as well. Thus, we expect a non-radiative loss of at least 0.6 eV per electron-hole recombination event from the hole trapping, plus potentially a contribution of similar magnitude from electron trapping in highly disordered material.   

\section{Conclusion}
\label{sec:concl}

In this work we examined the effect of cation disorder on the density and localization of electronic states in \ch{ZnGeN2}. The band gap of the system decreases significantly with decreasing degree of order from 3.5 eV for an ordered system to effectively 0 eV for strongly disordered systems. From non-dilute, disordered, but non-random structures with a significant degree of SRO, we calculated the DOS and the IPR of the material as a function of LRO and SRO, extracting band gaps as a function of both order parameters. We discussed the problem of defining the band gap in disordered materials and the ambiguities associated with the differentiation between defects and band states. While the topic deserves further discussion in the community, we used the IPR as a variable threshold for this separation. 

Localized, occupied states are caused by N with Zn-rich coordination. Isolated conduction band states attributed to cations are much less localized and occur independent of the cations' second shell coordination environment. Our findings in \ch{ZnGeN2} show a strong tendency for localized defect states to form at all order parameters other than the ground state, which could detrimentally impact carrier recombination in a \ch{ZnGeN2}-based device. This result indicates that SRO is important for inhibiting carrier localization, which corroborates recent findings in \ch{ZnSnN2:ZnO} with perfect SRO \cite{pan2020perfect}. Whereas in \ch{ZnSnN2:ZnO}, this perfect SRO phase can exist with long-range disorder, the direct relationship between SRO and LRO in \ch{ZnGeN2} means that LRO is a necessary, although not sufficient, requirement to minimize localization and non-radiative recombination in this system.

\section{Electronic structure calculation methods}
\label{sec:methods}

Data presented in this Article utilize the atomic configurations of Ref.~\citenum{cordell_2021_probing} to predict electronic structure properties as a function of the order parameter and effective temperature. In Figure \ref{fig:intro}, the electronic structure and density of states of \ch{ZnGeN2} were calculated in density functional theory (DFT) with band edge corrections to match the band gap from GW calculations (3.63 eV \cite{melamed2020combinatorial}) and plotted using pymatgen \cite{pymatgen}. To relax the lattice parameters, volume and ion positions of the 1,024 atom configurations from MC simulations, we used the generalized gradient approximation (GGA), Perdew-Burke-Ernzerhof (PBE) \cite{perdew_generalized_1997} type. Due to the large size of the supercells, a single k-point (1x1x1 mesh) sufficed for the relaxation using the gamma-point-only version of the Vienna Ab-initio Simulation Package (VASP). These calculations rely on Kresse-Joubert projector augmented wave datasets with pseudopotentials from VASP version 4.6 (i.e., Ge\_d, N\_s and Zn). The soft pseudopotential, N\_s, allows for a low energy cutoff of 380 eV that benefits the feasibility of calculations using large supercells. \cite{Kresse} Each supercell achieved convergence when the difference in energy between steps of the ionic relaxation dropped below $10^{-5}$ eV and forces below 0.02 eV $\si{\angstrom}^{-1}$ on each atom. These calculations used a Coulomb potential, $U - J = 6$ eV, applied to the Zn d orbital following the Dudarev approach \cite{Dudarev1998}.

The large size of the supercells precludes the possibility of applying the GW approach \cite{hedin1965new} for each structure. In place of GW methods, the DOS and IPR of relaxed structures were calculated using a parameterized single-shot hybrid functional with an additional Coulomb potential V (SSH+V) of -1.5 eV (comparable to a U parameter of +3 eV) applied to Zn d orbitals. \cite{Lany2017} The single-shot functional avoids the computationally expensive iteration to self-consistency of the hybrid functional Hamiltonian by holding the initial wavefunctions of the DFT+U calculation fixed.\cite{Lany2017} This non-self-consistent approach closely reproduces the GW electronic structure for Zn-IV-N$_2$ nitrides and nitride-oxide alloys \cite{Pan2018}. However, since the hybrid functional Hamiltonian depends on the band occupancies, ambiguities occur when the band gap incorrectly closes in the underlying DFT calculation. In this case, we perform a second SSH+V iteration with updated band occupancies. This extra step, which could introduce some additional uncertainty in the electronic structure, was needed for most 5,000 K configurations. The Hartree-Fock exchange mixing parameter of the SSH+V functional was set to 0.19 and screening parameter to 0 for all structures. These parameters were fitted to replicate the total DOS produced by GW calculations for the ground state structure with a band gap of 3.5 eV calculated in SSH+V. The same hybrid functional and V parameters were also used for the self-consistent calculation for non-radiative energy losses due to electron and hole trapping.

To be able to plot the DOS and IPR of various disordered configurations on a common energy axis (cf. Figure \ref{fig:dos}), a potential reference needs to be defined. The bare band energies (defined relative to the average electrostatic potential), are rather sensitive to changes of the cell volume. The volume of the disordered supercells increases by up to 0.9\% compared to the ground state for strongly disordered cells ($T_{eff}$ = 5,000 K) and by about 2\% for fully random cation disorder. To eliminate the shift of the band energies with cell volume, we performed a sequence of ordered \ch{ZnGeN2} calculations with varying cell volumes. Using the potential alignment approach of Ref.~\citenum{lany2009accurate} and linear regression, we obtained the potential shift $\Delta V_{pot}$ = -85 meV $\times$ $\Delta V_{vol}$, where $\Delta V_{vol}$ is the volume change in percent. This offset is subtracted before plotting the electronic structure in Figure \ref{fig:dos}.

\begin{acknowledgments}
This work was supported by the U.S. Department of Energy (DOE) under Contract No. DE-AC36-08GO28308 with the Alliance for Sustainable Energy, LLC, the manager and operator of the National Renewable Energy Laboratory. Funding was provided by the U.S. Department of Energy, Office of Energy Efficiency and Renewable Energy, Buildings Technologies Office. This work used high-performance computing resources located at NREL and sponsored by the Office of Energy Efficiency and Renewable Energy. The views expressed in the paper do not necessarily represent the views of the DOE or the U.S. government. The U.S. government retains and the publisher, by accepting the article for publication, acknowledges that the U. S. government retains a nonexclusive, paid-up, irrevocable, worldwide license to publish or reproduce the published form of this work, or allow others to do so, for government purposes.
\end{acknowledgments}

\section*{Data Availability Statement}
The data that support the findings of this study are available from the corresponding author upon reasonable request.

\bibliography{bibliography}

\end{document}